\documentclass[12pt]{article}
\usepackage[utf8]{inputenc}
\usepackage[comma,numbers,sort&compress]{natbib}
\usepackage{amsfonts,amsmath,amssymb,amsthm}
\usepackage{bm,amsbsy}
\usepackage{xcolor}
\usepackage{graphicx}
\usepackage[colorlinks=true, allcolors=blue]{hyperref}
\usepackage{mathtools}
\usepackage{sectsty} 

\usepackage[T1]{fontenc}
\usepackage[scaled]{helvet}

\usepackage{geometry}
\allowdisplaybreaks  

\newcommand{\asection}[1]{
	\renewcommand{\thesection}{Appendix \Alph{section}} \section{ #1}
        \addcontentsline{toc}{section}{Appendix \Alph{section}: #1}}

 \renewcommand{\eqref}[1]{(eq.~\!\ref{#1})}
      
\newcommand{\vx}{\mathbf{x}} 
\newcommand{\vw}{\mathbf{w}} 
\newcommand{\vy}{\mathbf{y}}
 
\newcommand{\vv}{\mathbf{v}}
\newcommand{\veps}{\bm{\epsilon}} 
\newcommand{\mA}{\mathbf{A}}
\newcommand{\mB}{\mathbf{B}}
\newcommand{\mK}{\mathbf{K}}
\newcommand{\mM}{\mathbf{M}}
\newcommand{\mN}{\mathbf{N}}
\newcommand{\mS}{\mathbf{S}}
\newcommand{\mV}{\mathbf{V}}

\newcommand{\mJ}{\mathbf{J}}
\newcommand{\mQ}{\mathbf{Q}}
\newcommand{\mI}{\mathbf{I}}
\newcommand{\mP}{\mathbf{P}}

\newcommand{\mR}{\mathbf{R}}
\newcommand{\mC}{\mathbf{C}}

\newcommand{\trp}{^{\top}}
\newcommand{\inv}{^{-1}}

\newcommand{\RR}{\mathbb{R}}

\newcommand{\vzero}{{\bm 0}}
\newcommand{\Nrm}{\mathcal{N}}

\title{Probing the relationship between linear dynamical systems and low-rank recurrent neural network models}
\usepackage{authblk}
\author[1]{Adrian Valente}
\author[1]{Srdjan Ostojic}
\author[2, 3]{Jonathan Pillow}
\affil[1]{Laboratoire de Neurosciences Cognitives et Computationnelles, INSERM U960, Ecole Normale Superieure - PSL Research University, 75005 Paris, France}
\affil[2]{Princeton Neuroscience Institute, Princeton University, Princeton, NJ, USA}
\affil[3]{Department of Psychology, Princeton University, Princeton, NJ, USA}
\date{October 2021}

\begin{document}

\maketitle


\begin{abstract}
\noindent A large body of work has suggested that neural populations  exhibit low-dimensional dynamics during behavior. However, there are a variety of different approaches for modeling low-dimensional neural population activity.  One approach involves latent linear dynamical system (LDS) models, in which population activity is described by a projection of low-dimensional latent variables with linear dynamics.  A second approach involves low-rank recurrent neural networks (RNNs), in which population activity arises directly from a low-dimensional projection of past activity. Although these two modeling approaches have strong similarities, they arise in different contexts and tend to have different domains of application. Here we examine the precise relationship between latent LDS models and linear low-rank RNNs.  When can one model class be converted to the other, and vice versa?  We show that latent LDS models can only be converted to RNNs in specific limit cases, due to the non-Markovian property of latent LDS models. Conversely, we show that linear RNNs can be mapped  onto LDS models, with latent dimensionality  at most twice the rank of the RNN.
\end{abstract}

\section{Introduction}

Recent work on large-scale neural population recordings has suggested that neural activity is often confined to a low-dimensional space, with fewer dimensions than the number of neurons in a population \cite{ChurchlandM07b,Gao15,Gallego2017,Saxena2019,Jazayeri-SO2021}.
To describe this activity, modellers have at their disposal a wide array of tools that give rise to different forms of low-dimensional activity \cite{Cunningham14}. Two  classes of modeling approaches that have generated a large following in the literature are: (1) descriptive statistical models; and (2) mechanistic models. Broadly speaking, descriptive statistical models aim to identify a probability distribution that captures the statistical properties of an observed neural dataset, while remaining agnostic about the mechanisms that gave rise to it.  Mechanistic models, by contrast, aim to reproduce certain characteristics of observed data using biologically-inspired mechanisms, but often with less attention to a full statistical description. Although these two classes of models often have similar mathematical underpinnings, there remain a variety of important gaps between them. 
Here we focus on reconciling the gaps between two simple but powerful  models of low-dimensional neural activity: latent linear dynamical systems (LDS) and low-rank linear recurrent neural networks (RNNs).  

The latent LDS model with Gaussian noise is a popular statistical model for low-dimen-sional neural activity in both systems neuroscience \cite{Smith03,Semedo14latent} and brain-machine interface settings \cite{Kim2008}.  This model has a long history in electrical engineering, where the problem of inferring latents from past observations has an analytical solution known as the Kalman filter \cite{Kalman60}.  In neuroscience settings, this model has been used to describe high-dimensional neural population activity in terms of  linear projections of low-dimensional latent variables with linear dynamics. Although this basic form of the model can only exhibit linear dynamics, recent extensions have produced state-of-the-art models for high-dimensional spike train data \cite{Yu06,Petreska11,Macke11nips,Pachitariu13,Archer14nips,Duncker19icml,Zoltowski20,Glaser20,Kim2008}.

Recurrent neural networks, by contrast, provide a powerful framework for building mechanistic models of neural population activity \citep{Sompolinsky88,laje2013,sussillo2014,rajan2016,barak2017}.  Although  randomly-connected RNN models typically have high-dimensional activity, recent work has shown that RNNs exhibit low dimensionality when constrained to have low-rank connectivity \cite{Mastrogiuseppe18}. Low-rank RNNs have been subsequently shown to be useful models able to reproduce many characteristics of low-dimensional neural trajectories \cite{landau2018,pereira2018,schuessler20,Beiran20, Dubreuil20,bondanelli2021}. Here we will focus on linear RNNs, which are less expressive but simpler to analyze than their non-linear counterparts, while still leading to rich dynamics \citep{hennequin14,kao21,bondanelli2021}. 

In this paper, we examine the mathematical relationship between latent LDS and low-rank linear RNN models. We show that even if both models produce Gaussian distributed activity patterns with low-dimensional linear dynamics, the two model classes have different statistical structure and are therefore not in general equivalent. More specifically, in latent LDS models, the output sequence has non-Markovian statistics, meaning that the activity in a single time step is not independent of its history given the activity on the previous time step.  This stands in contrast to linear RNNs, which are Markovian regardless of the rank of their connectivity.  A linear low-rank RNN can nevertheless provide a first-order approximation to the distribution over neural activity generated by a latent LDS model, and we show that this approximation becomes exact in several cases of interest, and in particular in the limit where the number of neurons is large compared to the latent dimensionality. Conversely, we show that any linear low-rank RNN can be converted to a latent LDS, although the dimensionality of the latent space depends on the overlap between the subspaces spanned by left and right singular vectors of the RNN connectivity matrix, and may be as high as twice the rank of this matrix. The two model classes are thus closely related, with linear low-rank RNNs comprising a subset of the broader class of latent LDS models.

\section{Modeling frameworks}
We start with a formal description of the two model classes in question, both of which describe the time-varying activity of a population of $n$ neurons.

\subsection{Latent LDS model} 
The latent linear dynamical system (LDS) model, also known as a linear-Gaussian state-space model, describes neural population activity as a noisy linear projection of a low-dimensional latent variable governed by linear dynamics with Gaussian noise \cite{Kalman60,Roweis99} (See schematic, Fig.~\ref{fig:schema}A). The model is characterized by the equations: \begin{alignat}{3}
  \label{eq:ldsx}
    \vx_{t+1} &=  \mA \vx_t \;  &+& \; \vw_t,  \qquad & \vw_t &\sim
    \Nrm(\vzero,\mQ) \\
      \label{eq:ldsy}
   \vy_{t} &= \mC \vx_t \;  &+ & \;  \vv_t, \qquad & \vv_t  &\sim \Nrm(\vzero,\mR).
 \end{alignat}
 Here, $\mathbf{x}_t$ is a $d$-dimensional latent (or "unobserved") vector that follows discrete-time linear dynamics specified by 
  a $d \times d$  matrix $\mA$, and is corrupted on each time step by a zero-mean Gaussian noise vector $\vw_t\in\RR^d$ with covariance $\mQ$. The vector of neural activity $\vy_{t}$ arises from a linear transformation of $\vx_{t}$ via the
  $n \times d$ observation (or ``emissions'') matrix $\mC$, corrupted by zero-mean Gaussian noise vector $\vv_t\in \RR^n$ with covariance $\mR$. Generally we assume $d<n$, so
 that the high-dimensional observations $\vy_t$ are explained by the
 lower-dimensional dynamics of the latent vector $\vx_t$.
 
The complete model also contains a specification of the distribution of the initial latent vector $\vx_0$, which is commmonly assumed to have a zero-mean Gaussian distribution  with covariance $\bm\Sigma_0$:
 \begin{equation}
   \label{eq:2}
   \vx_0 \sim \Nrm(\vzero,\bm\Sigma_0).
 \end{equation}
The complete parameters of the model are thus $\theta_{LDS} = \{ \mA,\mC,\mQ,\mR,\bm\Sigma_0\}$.  Note that this parametrization of an LDS is not unique: any invertible linear transformation of the latent space leads to an equivalent model if the appropriate transformations are applied to matrices $\mA$, $\mC$, $\mQ$, and $\bm\Sigma_0$.

\subsection{Low-Rank Linear RNN}

\begin{figure}
    \centering
    \includegraphics[width=\textwidth]{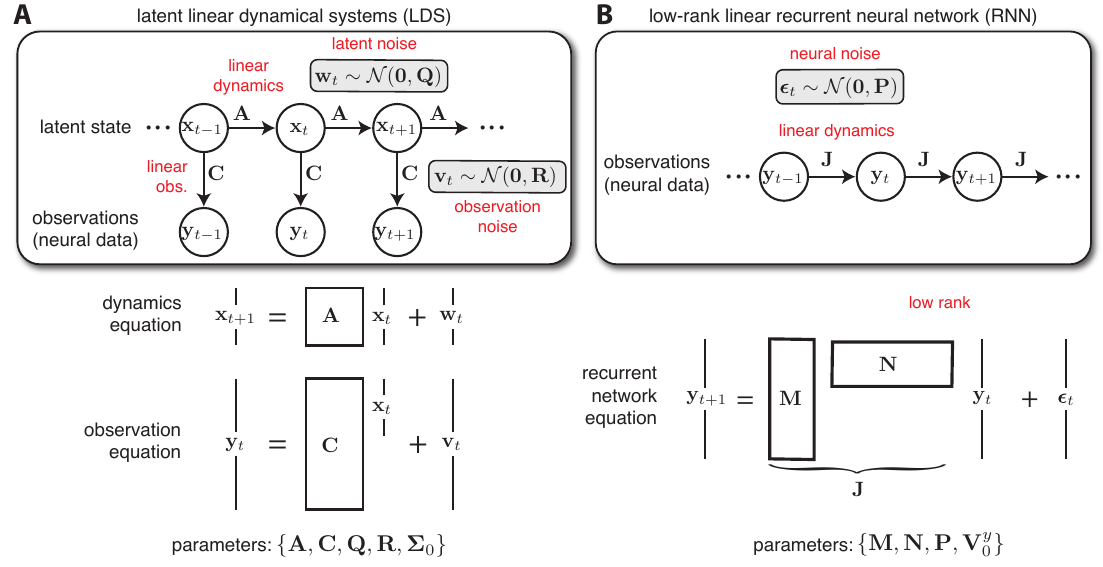}
    \caption{\textbf{(A)} Schematic representation of the latent linear dynamical system model, as defined by (eqs.~\ref{eq:ldsx}-\ref{eq:2}). \textbf{(B)} Schematic representation of the low-rank linear RNN, as defined by (eqs.~\ref{eq:rnn}-\ref{eq:rnn_ic}).}
    \label{fig:schema}
\end{figure}

A linear RNN, also known as an auto-regressive (AR) model, represents observed neural activity as a noisy linear projection of the activity at the previous timestep.  We can write the model as (Fig. \ref{fig:schema}B):
\begin{equation}
  \label{eq:rnn}
  \vy_{t+1} = \mJ \vy_t + \veps_t, \qquad \veps_t \sim \Nrm(\vzero,\mP),
\end{equation}
where  $\mJ$ is an $n \times n$ recurrent weight matrix,
and $\veps_t\in \RR^n$ is a Gaussian noise vector with mean zero and
covariance $\mP$. We moreover assume that the initial condition is drawn from a zero-mean distribution with covariance $\mV_0^y$:

\begin{equation}
   \label{eq:rnn_ic}
   \vy_0 \sim \Nrm(\vzero,\mV_0^y).
 \end{equation}
 
A low-rank RNN model is obtained by  constraining the rank of the recurrent weight matrix $\mJ$ to be $r \ll n$. In this case the recurrence matrix can be factorized as 
\begin{equation}
\label{eq:low_rank}
    \mJ = \mM \mN \trp,
\end{equation}
where $\mM$ and $\mN$ are both $n \times r$ matrices. 

Note that this factorization is not unique, but a particular
factorization can be obtained from a low-rank $\mJ$ matrix using the
truncated singular value decomposition: $\mJ
=\mathbf{U}\mathbf{S}\mV\trp$, where $\mathbf{U}$ and $\mathbf{V}$ are
semi-orthogonal $n\times r$ matrices of left and right singular
vectors, respectively, and $\mathbf{S}$ is an $r \times r$ diagonal matrix containing the largest
singular values.  We can then set $\mM = \mathbf{U}$ and $\mN = \mS \mV\trp$. 


The model parameters of the low-rank linear RNN are therefore given by $\theta_{RNN} = \{\mM, \mN, \mP, \mV_0^y\}$.  

\subsection{Comparing the two models}

Both models described above exhibit low-dimensional dynamics embedded in a high-dimensional observation space. In the following, we examine the probability distributions $P(\vy_1, \ldots, \vy_T)$ over time series $(\vy_1, \ldots, \vy_T)$ generated by the two models. We show that in general, the two models give rise to different distributions, such that the family of probability distributions generated by the LDS model cannot all be captured with low-rank linear RNNs.  Specifically, RNN models are constrained to purely Markovian distributions,  which is not the case for LDS models. However, the two model classes can be shown to be equivalent when the observations $\vy_t$ contain exact information about the latent state $\vx_t$, which is in particular the case if the observation noise is orthogonal to the latent subspace, or in the limit of a large number of neurons $n \gg d$. Conversely, a low-rank linear RNN can in general be mapped to a latent LDS with a dimensionality of the latent state at most twice the rank of the RNN. 

\section{Mapping from LDS models to linear low-rank RNNs}

\subsection{Non-equivalence in the general case}
Let us consider an LDS described by (eqs.~\ref{eq:ldsx}-\ref{eq:2}) and  a low-rank linear RNN defined by (eqs.~\ref{eq:rnn}-\ref{eq:rnn_ic}).
We start by comparing the properties of the joint distribution $P(\vy_0, \dots, \vy_T)$ for any value of $T$ for the two models. For both models, the joint distribution can be factored under the form:
\begin{equation}
\label{eq:factdist}
   P(\vy_0, \dots, \vy_T) = P(\vy_0)\prod_{t=1}^{T} P(\vy_t \mid \vy_{t- 1}, \dots, \vy_0),
\end{equation}
where each term in the product is the distribution of neural population activity at a single time point given all previous activity  (see \ref{sec:appA} for details). More specifically, each of the conditional distributions in \eqref{eq:factdist} is Gaussian, and for the LDS we can parametrize these distributions as:
\begin{align}
    \label{eq:recnormyx}
    P(\vx_{t} | \vy_{t-1}, \dots, \vy_0) &:= \Nrm(\hat{\vx}_{t}, \mV_{t})\\
    P(\vy_{t} | \vy_{t-1}, \dots, \vy_0) &= \Nrm(\mC\hat{\vx}_{t}, \mC\mV_{t}\mC\trp + \mR),
    \label{eq:recnormxy}
\end{align}
where $\hat{\vx}_{t}$ is the mean of the conditional distribution over the latent at timestep $t$, given observations until timestep $t-1$.  It obeys the recurrence equation:
\begin{align}
\label{eq:recxhat}
    \hat{\vx}_{t+1} &= \mA(\hat{\vx}_{t} + \mK_{t}(\vy_t - \mC\hat{\vx}_{t})),
\end{align}
where $\mK_t$ is the Kalman gain given by
\begin{equation}
\label{eq:kalgain}
\mK_t = \mV_t\mC\trp(\mC\mV_t\mC\trp + \mR)\inv.
\end{equation}
and $\mV_{t}$ represents a covariance matrix, which is independent of the observations and follows a recurrence equation detailed in \ref{sec:appA}.

Iterating equation (eq.~\ref{eq:recxhat}) over multiple timesteps, one can see that $\hat{\vx}_{t+1}$ depends not only on the last observation $\vy_t$, but on the full history of observations $(\vy_0, \ldots, \vy_t)$, which therefore affects the distribution at any given timestep. The  process $(\vy_0, \ldots, \vy_t)$ generated by the LDS model is hence non-Markovian.

Conversely, for the linear RNN, the observations $(\vy_0, \ldots, \vy_t)$ instead {\it do} form a Markov process, meaning that observations are conditionally independent of their history given the activity from the previous timestep: \begin{equation}
    P(\vy_t \mid  \vy_{t-1}, \ldots , \vy_0) = P(\vy_t \mid  \vy_{t-1}).
\end{equation}
The fact that this property does not in general hold for the latent LDS shows that the two model classes are not equivalent. Due to this fundamental constraint, the RNN can only approximate the complex distribution (eq.~\ref{eq:factdist}) parametrized by an LDS, as detailed in the following section and illustrated in figure \ref{fig:autocov}.

\subsection{Matching the first-order marginals of an LDS model}
\label{mapping-lds-rnn-approx}

We can obtain a Markovian approximation of the LDS-generated sequence of observations $(\vy_0, \ldots, \vy_t)$ by deriving the conditional
distribution $P(\vy_{t+1} \mid \vy_t)$ under the LDS model, and matching it with a low-rank RNN \citep{Pachitariu13}. This type of first-order approximation will preserve exactly the one-timestep-difference marginal distributions $P(\vy_{t+1}, \vy_t)$ although structure across longer timescales might not be captured correctly. 

First, let us note that we can express both $\vy_{t}$ and $\vy_{t+1}$ as  noisy linear
projections of~$\vx_{t}$:
\begin{alignat}{3}
   \vy_{t} &=  \mC \vx_{t} \;  &+& \; \vv_{t}, \\
   \vy_{t+1} &= \mC (\mA \vx_{t} + \vw_t) \;  &+ & \;  \vv_{t+1}, 
 \end{alignat}
 which follows from (eq.~\ref{eq:ldsx}).
 
 Let $\Nrm(\vzero, \bm\Sigma_{t})$ denote the Gaussian marginal distribution
 over the latent vector $\vx_{t}$ at time $t$.  Then we can use
 standard identities for linear transformations of Gaussian
 variables to derive the joint distribution over $\vy_t$ and $\vy_{t+1}$:
\begin{equation}
  \begin{bmatrix}
    \vy_{t} \\ \vy_{t+1}
  \end{bmatrix} \sim \Nrm \left(
    \begin{bmatrix}
      \vzero \\ \vzero
    \end{bmatrix},
    \begin{bmatrix}
      \mC\bm\Sigma_t\mC\trp + \mR &  \mC\bm\Sigma_t\mA\trp \mC\trp
      \\ \mC\mA\bm\Sigma_t \mC\trp &
      \mC(\mA\bm\Sigma_t\mA\trp + \mQ)\mC\trp + \mR
    \end{bmatrix}
    \right).
  \end{equation}
 We can then apply the formula for the conditionalization of Gaussians (see \cite{Bishop} equations (2.81) - (2.82)) to
 obtain:
 \begin{equation}
   \label{eq:1}
   \vy_{t+1} \mid \vy_t \sim \Nrm \left(\mJ_t \vy_t, \mP_t  \right)
 \end{equation}
 where
 \begin{align}
   \label{eq:ldstornn}
   \mJ_t &=  \mC\mA\bm\Sigma_t  \mC\trp
         (\mC\bm\Sigma_t\mC\trp + \mR)\inv \\
    \label{eq:ldstornnvar}
   \mP_t &=  \mC(\mA\bm\Sigma_t\mA\trp + \mQ)\mC\trp + \mR - \mC\mA\bm\Sigma_t \mC\trp  (\mC\bm\Sigma_t\mC\trp + \mR)\inv \mC\bm\Sigma_t\mA\trp \mC\trp.
 \end{align}
 
In contrast, from (eq.~\ref{eq:rnn}), for a low-rank RNN the first-order marginal is given by
\begin{equation}\label{eq:marginal_rnn}
   \vy_{t+1} \mid \vy_t \sim \Nrm \left(\mJ \vy_t, \mP  \right).
\end{equation}

Comparing equations \eqref{eq:1} and \eqref{eq:marginal_rnn}, we see  for the LDS model, the effective weights $\mJ_t$ and the covariance $\mP_t$  depend on time through $\bm\Sigma_t$, the marginal covariance of the latent at time $t$, while for the RNN they do not. 
Note however that $\bm\Sigma_t$ follows the recurrence relation
\begin{equation}
    \bm\Sigma_{t+1} = \mA \bm\Sigma_t \mA \trp + \mQ
\end{equation}
which converges towards a fixed point $\bm\Sigma_\infty$ that obeys the discrete Lyapunov equation
\begin{equation}
\label{eq:riccati}
\bm\Sigma_\infty = \mA \bm\Sigma_\infty \mA\trp + \mQ,
\end{equation}
provided all eigenvalues of $\mA$ have absolute value less than 1. 
 
The LDS can therefore be approximated by an RNN with constant weights
when the initial covariance $\bm\Sigma_0$ is equal to the asymptotic
covariance $\bm\Sigma_\infty$, as noted previously
\cite{Pachitariu13}. Note that even if this condition does not hold at
time 0, $\bm\Sigma_\infty$ will in general be a good approximation of
the latent covariance after an initial transient. In this case we
obtain the fixed recurrence weights: 
\begin{equation}
\label{eq:ldstornnJ}
    \mJ =  \mC\mA\bm\Sigma_\infty  \mC\trp (\mC\bm\Sigma_\infty \mC\trp + \mR)\inv := \mM\mN\trp
\end{equation}
where we define $\mM = \mC$ which has shape $n \times d$ and $\mN\trp = \mA\bm\Sigma_\infty  \mC\trp (\mC\bm\Sigma_\infty \mC\trp + \mR)\inv$ which has shape $d \times n$, so that $\mJ$ is a rank $r$ matrix with $r=d$.

\subsection{Cases of equivalence between LDS and RNN models}
Although latent LDS and low-rank linear RNN models are not equivalent
in general, we can show that the first-order Markovian approximation
introduced above becomes exact in two limit  cases of interest: (i)
for observation noise orthogonal to the latent subspace, and (ii) in
the limit $n \gg d$, with coefficients of the observation matrix
generated randomly and    independently.  

Our key observation is that if $\mathbf{K}_t\mC = \mI$ in
 \eqref{eq:recxhat} with $\mI$ the identity matrix, we have $\hat{\vx}_{t+1} = \mA
\mK_t\vy_t$, so that the dependence on the observations before
timestep $t$ disappears, and the LDS therefore becomes Markovian.
Interestingly, this condition $\mK_t\mC = \mI$ also
implies that the latent state can  be inferred from the
current observation $\vy_t$ alone (see \eqref{eq:kalmanxstep2}) and
that this inference is exact, since the variance of the distribution
$p(\vx_t | \vy_t)$ is equal to 0 as seen from
\eqref{eq:kalmanVstep2}. We next examine two cases where this
condition is satisfied. 

We first consider the situation where the observation noise vanishes, ie. $\mR = \boldsymbol{0}$. Then, as shown in \ref{sec:appA}, the Kalman gain is $\mK_t = (\mC\trp\mC)\inv\mC\trp$, so that $\mK_t\mC = \mI$. In that case, the approximation of the LDS by the RNN defined in the section 3.2 is exact, with equations \eqref{eq:ldstornn} and \eqref{eq:ldstornnvar} becoming:
\begin{align}
    \mJ &= \mC\mA(\mC\trp\mC)\inv\mC\trp \\
    \label{eq:pnonoise}
    \mP &= \mC\mQ\mC\trp.
\end{align}
More generally, this result remains valid when the observation noise
is orthogonal to the latent subspace spanned by the columns of the
observation matrix $\mC$ (in which case the recurrence noise given by \eqref{eq:pnonoise}
becomes $\mP = \mC\mQ\mC\trp + \mR$).

A second case in which we can obtain $\mK_t\mC \approx \mI$ is  in the
limit of many neurons, $n \gg d$,  assuming that coefficients of the observation matrix are generated randomly and independently. Indeed, under these hypotheses the Kalman gain given by equation \eqref{eq:kalgain} is dominated by the  term $\mC\mV_t\mC\trp$, so that the observation covariance $\mR$ becomes negligible, as shown formally in \ref{sec:appB}. Intuitively this means that the information about the latent state $\hat{\vx}_t$ is distributed over a large enough population of neurons for the Kalman filter to average out the observation noise and estimate it optimally without making use of previous observations.
 Ultimately, this makes the LDS asymptotically Markovian in the case
 where we have an arbitrarily large neural population relative to the
 number of latent dimensions.





\begin{figure}[t]
    \centering
    \includegraphics[width=0.99\textwidth]{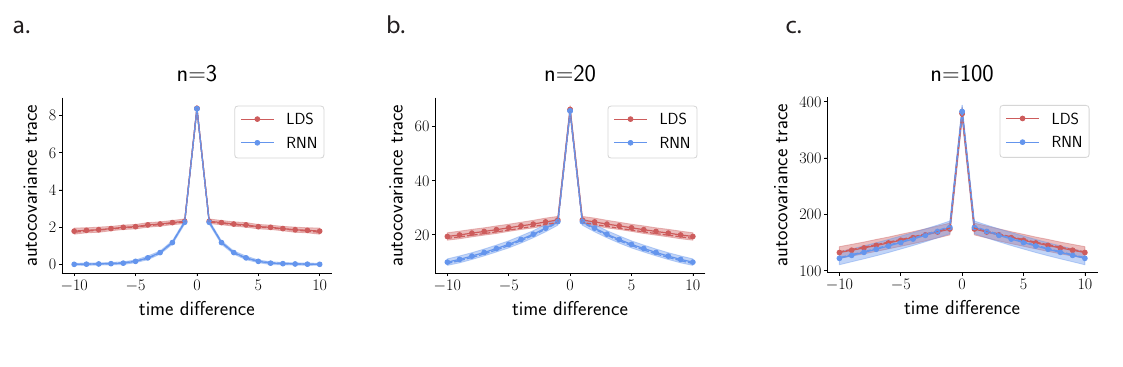}
    \caption{Trace of the autocovariance matrix of observations $\mathbf{y}_t$ for example LDS models compared with their first-order RNN approximations. The latent space is one-dimensional ($d=1$), and the dimension $n$ of the observation space is increased from left to right: a. $n=3$, b. $n=20$, c. $n=100$.
     The parameters of the latent state processes are fixed scalars ($\mathbf{A}=(0.97)$, $\mathbf{Q}=(0.1)$), while the elements of the observation matrices $\mathbf{C}$ are drawn randomly and independently from a centered Gaussian distribution of variance 1. The observation noise has covariance $\mR = \sigma_v^2\mI_n$ with $\sigma_v^2 = 2$. Note that we have chosen observation noise to largely dominate over latent state noise in order to obtain a large difference between models at low $n$. Dots and shaded areas indicate respectively mean and standard deviation of different estimations of the autocovariance trace run on 10 independent folds of 100 trials each (where $\mC$ was identical across trials).
    }
    \label{fig:autocov}
\end{figure}

To illustrate the convergence of the low-rank RNN approximation to the
target LDS in the large $n$ limit, in figure \ref{fig:autocov} we
consider a simple example with a one-dimensional latent space and
observation spaces of increasing dimensionality. To visualize the
difference between the LDS and its low-rank RNN approximation, we plot
the trace of the autocovariance matrix of observations $\vy_t$ in the
stationary regime,
$\rho(\delta) = \operatorname{Tr}(\mathbb{E}[\vy_t\vy_{t+\delta}^T])$.
Since the RNNs are constructed to capture the marginal distributions
of observations separated by at most one timestep, the two curves
match exactly for a lag $\delta\in\{-1, 0, 1\}$, but dependencies at
longer timescales cannot be accurately captured by an RNN due to its
Markov property (Fig. \ref{fig:autocov}a). However, these
differences vanish as the dimensionality of the observation space
becomes much larger than that of the latent space (Fig.
\ref{fig:autocov}b-c), which illustrates that the LDS converges to a process equivalent to a
low-rank RNN.

\section{Mapping  low-rank linear RNNs onto LDS models}
\label{sec:rnntolds}
We now turn to the reverse question: under what conditions can a
low-rank linear RNN be expressed as a linear LDS model? We start with
an intuitive mapping for the deterministic case (\emph{i.e.}, when
noise covariance $\mP = \vzero$), and then extend it to a more general mapping valid in
the presence of noise.

We first consider a deterministic linear low-rank RNN obeying:
\begin{equation}
    \vy_{t+1} = \mM\mN\trp\vy_t.
\end{equation}
Since $\mM$ is an $n\times r$ matrix, it is immediately apparent that for all $t$, $\vy_t$ is confined to a linear subspace of dimension $r$, spanned by the columns of $\mM$.
Hence, we can define the latent state as
\begin{equation}
\vx_t = \mM^\#\vy_t
\end{equation}
where $\mM^\#$ is the pseudoinverse of $\mM$ defined as $\mM^\# = (\mM\trp\mM)\inv\mM\trp$, so that we  retrieve $\vy_t$ as:
\begin{equation}
\label{eq:ymx}
    \vy_t = \mM\vx_t.
\end{equation}
We  then obtain a recurrence equation for the latent state :
\begin{align} \label{eq:rnn_latent_dyn}
    \vx_{t+1} &= \mM^\# \vy_{t+1} \nonumber \\
              &= \mM^\#\mM\mN\trp\vy_t \nonumber \\
              &=\mN\trp\mM\vx_t \nonumber \\
              &\coloneqq \mA\vx_t
\end{align}
which describes the dynamics of a deterministic linear LDS with $\mA = \mN\trp\mM$. A key insight from \eqref{eq:rnn_latent_dyn} is that  the latent dynamics project the activity from the previous timestep onto the column space of $\mN$, which therefore determines the part of the activity that is integrated by the recurrent dynamics \citep{Mastrogiuseppe18, schuessler20, Beiran20, Dubreuil20}.

In presence of noise $\veps_t$ in the RNN dynamics, $\vy_t$ is no longer confined to the column space of $\mM$. Part of the additional activity can be represented as observation noise that is independent across time steps. Another part however stems from RNN noise integrated from the previous timesteps. As noted above, recurrent dynamics only integrate the activity in the column space of $\mN$. In presence of noise, this part of state space therefore needs to be included into the latent variables. Note that a similar observation can be made about external inputs when they are added to the RNN dynamics (see \ref{sec:appD}).


A full mapping from a noisy low-rank RNN to an LDS model can therefore be built by extending the latent space to the linear subspace $\mathcal{F}$ of $\RR^n$ spanned by the columns of $\mM$ and $\mN$ (see \ref{sec:appC}), which has dimension $d$ with $k \leq d \leq 2k$. Let $\mC$ be a matrix whose columns form an orthogonal basis for this subspace (which can be obtained via the Gram-Schmidt algorithm). In that case we can define the latent vector as : 
\begin{equation}
    \vx_t = \mC\trp \vy_t,
\end{equation}
and the latent dynamics are given by
\begin{equation}
    \vx_{t+1} = \mathbf{A}\vx_t + \mathbf{w}_t,
\end{equation}
\begin{sloppypar}
where  the recurrence matrix is $\mA = \mC\trp\mJ\mC$, and the latent
dynamics noise is ${\mathbf{w}_t \sim \Nrm(\vzero, \mQ)}$ with $\mQ =
\mC\trp\mP\mC$. Introducing $\vv_t = \vy_t -
\mC\vx_t$, under a specific condition on the noise covariance $\mP$ we
obtain a normal random variable independent of the other sources of
noise in the process (\ref{sec:appC}), so that $\vy_t$ can be
described as a noisy observation of the latent state $\vx_t$ as in the
LDS model.
\end{sloppypar}


\section{Discussion}
In this note we have examined the relationship between two  simple yet powerful classes of models of low-dimensional activity: linear latent dynamical systems (LDS) and  low-rank linear recurrent neural networks (RNN). We have focused on these tractable linear models with additive Gaussian noise to highlight their mathematical similarities and differences. Although both models induce a jointly Gaussian distribution over neural population activity, generic latent LDS models can exhibit long-range, non-Markovian temporal dependencies that cannot be captured by low-rank linear RNNs, which describe neural population activity with a first-order Markov process. Conversely, we showed that generic low-rank linear RNNs can be captured by an equivalent latent LDS model.
However, we have shown that the two classes of models are effectively equivalent in limit cases of practical interest for neuroscience, in particular when the number of sampled neurons is much higher than the latent dimensionality. 

Although these two model classes can generate similar sets of neural trajectories, different approaches are typically used for fitting them to neural data: parameters of  LDS models are in general inferred by variants of the expectation-maximization algorithm \citep{Yu06,Pachitariu13,nonnenmacher2017, durstewitz2017}, which include the Kalman smoothing equations \cite{Roweis99}, while RNNs are often fitted with variants of linear regression \cite{rajan2016,Eliasmith03, Pollock20, Bondanelli21} or backpropagation-through-time \cite{Dubreuil20}. The relationship uncovered here therefore opens the door to comparing different fitting approaches  more directly, and in particular to developing probabilistic methods for inferring RNN parameters from data.

We have considered here only linear RNN and LDS models. Non-linear low-rank RNNs without noise can be directly reduced to non-linear latent dynamics with linear observations following the same mapping as in Section 4 \citep{Mastrogiuseppe18, schuessler20, Beiran20, Dubreuil20}, and therefore define a natural class of non-linear LDS models. A variety of other non-linear generalizations of LDS models have been considered in the litterature. One line of work has examined linear latent dynamics with a non-linear observation model \cite{Yu06} or non-linear latent dynamics \cite{Yu06, durstewitz2017, Duncker19icml, Pandarinath18, Kim2008}. Another line of work has focused on switching LDS models \cite{Linderman17, Glaser20} for which the system undergoes different linear dynamics depending on a hidden discrete state, thus combining elements of latent LDS and hidden Markov models. Both non-linear low-rank RNNs and switching LDS models are universal approximators of low-dimensional dynamical systems \citep{Funahashi93approximation, Chow2000, Beiran20}.  Relating switching LDS models to local linear approximations of non-linear low-rank RNNs \citep{Beiran20, Dubreuil20} is therefore an interesting avenue for future investigations.

\section*{Acknowledgements}
AV and SO were supported by the program “Ecoles Universitaires de Recherche” ANR-17-EURE-0017, the CNCRS program through French Agence Nationale de la Recherche (ANR-19-NEUC-0001-01) and the NIH BRAIN initiative (U01NS122123). JWP was supported by grants from the Simons Collaboration on the Global Brain (SCGB AWD543027), the NIH BRAIN initiative (R01EB026946), and by a visiting professorship grant from the Ecole Normale Superieure (ENS).

\bibliographystyle{unsrtnat} 
\bibliography{LDSvRNNbib.bib}  

\clearpage

\setcounter{section}{0}

\asection{Kalman filtering equations} \label{sec:appA}
We reproduce in this appendix the recurrence equations followed by the conditional distributions in equation \eqref{eq:factdist} for both the latent LDS and the linear RNN models.

For the latent LDS model, the conditional distributions are Gaussians and their form is given by the Kalman filter equations \cite{Kalman60,BYutechreport04, Welling10}. Following \cite{BYutechreport04}, we observe that for any two timesteps $\tau \leq t$ the conditional distributions $p(\vy_{t+1} | \vy_\tau, \dots, \vy_0)$ and $P(\vx_{t+1} | \vy_\tau, \dots, \vy_0)$ are Gaussian, and we introduce the notations :
\begin{align}
\label{eq:recnormy2}
    P(\vy_{t} | \vy_\tau, \dots, \vy_0) &:= \Nrm(\hat{\vy}_{t}^\tau, \mathbf{W}_{t}^\tau) \\
    P(\vx_{t} | \vy_\tau, \dots, \vy_0) &:= \Nrm(\hat{\vx}_{t}^\tau, \mathbf{V}_{t}^\tau) 
    \label{eq:recnormx2}
\end{align}
In particular, we are interested in expressing $\hat{\vy}_{t+1}^t$ and $\hat{\vx}_{t+1}^t$, which are the predicted future observation and latent state, but also in $\hat{\vx}_t^t$ which represents the  latent state inferred from the history of observations until timestep $t$ included. To lighten notations, in the main text we remove the exponent when it has one timestep difference with the index, by writing $\hat{\vx}_{t+1}$, $\hat{\vy}_{t+1}$, $\mathbf{W}_{t+1}$ and $\mathbf{V}_{t+1}$ instead of respectively $\hat{\vx}_{t+1}^t$, $\hat{\vy}_{t+1}^t$, $\mathbf{W}_{t+1}^t$ and $\mathbf{V}_{t+1}^t$.

First, note that we have the natural relationships:

\begin{align}
    \label{eq:kalmanxstep1}
     \hat{\vx}_{t+1}^t &= \mA \hat{\vx}_t^t \\
     \hat{\vy}_{t+1}^t &= \mC \hat{\vx}_{t+1}^t \\
    \label{eq:kalmanVstep1}
     \mathbf{V}_{t+1}^t &= \mA \mathbf{V}_t^t \mA\trp + \mQ \\
     \mathbf{W}_{t+1}^t &= \mC \mathbf{V}_{t+1}^t \mC\trp + \mR
\end{align}
so that it is sufficient to find expressions for $\hat{\vx}_t^t$ and $\mathbf{V}_t^t$. After calculations detailed in \cite{BYutechreport04} or \cite{Welling10}, we obtain:

\begin{align}
    \label{eq:kalmanxstep2}
    \hat{\vx}_{t}^{t} &= \hat{\vx}_{t}^{t-1} + \mathbf{K}_t(\vy_t - \mC\hat{\vx}_{t}^{t-1}) \\
    \label{eq:kalmanVstep2}
    \mathbf{V}_{t}^{t} &= (\mI - \mathbf{K}_t\mC)\mathbf{V}_{t}^{t-1}
\end{align}
where $\mathbf{K}_t$ is the Kalman gain given by:
\begin{equation}
\label{eq:kalgain2}
\mathbf{K}_t = \mathbf{V}^{t-1}_t\mC\trp(\mC\mathbf{V}^{t-1}_t\mC\trp + \mR)\inv.
\end{equation}

These equations form a closed recurrent system, as can be seen by combining \eqref{eq:kalmanxstep1} and \eqref{eq:kalmanxstep2}, and \eqref{eq:kalmanVstep1} and \eqref{eq:kalmanVstep2} to obtain a self-consistent set of recurrence equations for the predicted latent state and its variance:
\begin{align}
    \label{eq:kalmanxsimple}
    \hat{\vx}_{t+1}^t &= \mA(\hat{\vx}_t^{t-1} + \mathbf{K}_t(\vy_t - \mC\hat{\vx}_t^{t-1})) \\
    \label{eq:kalmanVsimple}
    \begin{split}
    \mathbf{V}^{t}_{t+1} &= \mA(\mI - \mathbf{K}_t\mC)\mathbf{V}^{t-1}_t \mA\trp + \mQ \\
    &= \mA (\mI - \mathbf{V}^{t-1}_t\mC\trp(\mC\mathbf{V}^{t-1}_t\mC\trp + \mR)\inv\mC)\mathbf{V}^{t-1}_t\mA\trp + \mQ 
    \end{split}
\end{align}

From \eqref{eq:kalmanxsimple} we see that the predicted state at time $t+1$, and thus the predicted observation, depends on observations at time steps $\tau \leq t-1$ through the term $\hat{\vx}_t$, making the system non-Markovian. Also note that equations for the variances don't involve any of the observations $\vy_t$, showing these are exact values and not estimations.

This derivation however is not valid in the limit case $\mR = \boldsymbol{0}$, since $\mathbf{K}_t$ is then undefined. In that case however, we can observe that $\vy_t$ lies in the linear subspace spanned by the columns of $\mC$, so that one can simply replace \eqref{eq:kalmanxstep2} by:
\begin{equation}
\hat{\vx}_{t}^t = \mathbf{C}^\# \vy_t = \vx_t,
\end{equation}
where $\mathbf{C}^\# = (\mC\trp\mC)\inv\mC\trp$ is the pseudoinverse of $\mC$. Since this equation is deterministic, the variance of the estimated latent state is equal to $\boldsymbol{0}$, so that equation \eqref{eq:kalmanVstep2} becomes $\mathbf{V}_t^t = \boldsymbol{0}$. This case can be encompassed by equations \eqref{eq:kalmanxstep1} - \eqref{eq:kalmanVstep2} if we rewrite the Kalman gain as:
\begin{equation}
    \mathbf{K}_t = \mathbf{C}^\# = (\mC\trp\mC)\inv\mC\trp.
\end{equation}

Finally, for the linear RNN, the conditional distribution of equation \eqref{eq:recnormy2} is directly given by :
\begin{align}
    P(\vy_{t+1} | \vy_t, \dots, \vy_0) &= \Nrm(\mJ\vy_t, \mP)
\end{align}
which shows that the predicted observation only depends on the last one, making the system Markovian.

\asection{Equivalence in the large network limit}
\label{sec:appB}
Here we make the assumption that the coefficients of the observation matrix are generated randomly and independently. We show that in the  limit of large $n$ with $d$ fixed one obtains $\mathbf{K}_t\mC \to \mI$ so that the LDS is asymptotically Markovian and can therefore be exactly mapped to an RNN. \\
 
 We start by considering a linear LDS whose conditional distributions obey equations \eqref{eq:recnormy2} - \eqref{eq:kalmanVsimple}, with the Kalman gain obeying \eqref{eq:kalgain2}. To simplify \eqref{eq:kalgain2}, we focus on the steady state  where variance $\mathbf{V}_t$ has reached its stationary limit $\mathbf{V}$ in \eqref{eq:kalmanVsimple}. 
 
 Without loss of generality, we reparametrize the LDS by applying a change of basis to the latent states such that $\mathbf{V} = \mI$. We also apply a change of basis to  the observation space such that $\mR = \mI$ in the new basis (this transformation does not impact the conditional dependencies between the $\vy_t$ at different timesteps, and it can also be shown that it cancels out in the expression $\mathbf{K}_t\mC$). The equation \eqref{eq:kalgain2} then becomes: 
$$\mathbf{K}_t\mC = \mC\trp(\mI + \mC\mC\trp)\inv\mC.$$
Applying the matrix inversion lemma gives $(\mI + \mC\mC\trp)\inv = \mI - \mC(\mI + \mC\trp\mC)\inv\mC\trp$, from which we get:
$$\mathbf{K}_t\mC = \mC\trp\mC - \mC\trp\mC(\mI + \mC\trp\mC)\inv\mC\trp\mC.$$
Using a Taylor expansion we then write:
\begin{align*}
    (\mI + \mC\trp\mC)\inv &= (\mI + (\mC\trp\mC)\inv)\inv(\mC\trp\mC)\inv\\
    &= (\sum_{k=0}^\infty (-(\mC\trp\mC)\inv)^k)(\mC\trp\mC)\inv \\
    &\approx (\mC\trp\mC)\inv - ((\mC\trp\mC)\inv)^2 + ((\mC\trp\mC)\inv) ^3,
\end{align*}

which gives:
\begin{align*}
    \mathbf{K}_t\mC &\approx \mC\trp\mC - \mC\trp\mC(\mC\trp\mC)\inv\mC\trp\mC + \mC\trp\mC((\mC\trp\mC)\inv)^2\mC\trp\mC - \mC\trp\mC((\mC\trp\mC)\inv)^3\mC\trp\mC \\
    & \approx \mC\trp\mC - \mC\trp\mC + \mI - (\mC\trp\mC)\inv.
 \end{align*}

Assuming  the coefficients of the observation matrix  are iid. with zero mean and unit variance, 
for $n$ large we obtain $\mC\trp\mC = n\mI + \mathcal{O}(\sqrt{n})$ from the central limit theorem, so that $(\mC\trp\mC)\inv = \mathcal{O}(1/n)$ (which can again be proven with a Taylor expansion). This finally leads to $\mathbf{K}_t \mC = \mI + \mathcal{O}(1/n)$.

\asection{Derivation of the RNN to LDS mapping}
\label{sec:appC}
As mentioned in section \ref{sec:rnntolds}, we consider an RNN defined by \eqref{eq:rnn}, with $\mJ = \mM\mN\trp$ and note $\mC$ an orthonormal matrix whose columns form a basis of $\mathcal{F}$, the linear subspace spanned by the columns of $\mM$ and $\mN$.
Note that $\mC\mC\trp$ is an orthogonal projector onto the subspace $\mathcal{F}$, and that since all columns of $\mM$ and $\mN$ belong to this subspace we have $\mC\mC\trp\mM = \mM$ and $\mC\mC\trp\mN = \mN$. Hence, we have:
\begin{equation}\label{eq:eqJ}
\mC\mC\trp\mJ\mC\mC\trp = \mJ.
\end{equation}

We thus define the latent vector as $ \vx_t = \mC\trp \vy_t$, and we can then write:
\begin{align*}
    \vx_{t+1} &= \mC\trp \vy_{t+1} \\
              &= \mC\trp \mJ \vy_t + \mC\trp \veps_t \\
              &= \mC\trp \mC\mC\trp\mJ\mC\mC\trp \vy_t + \mC\trp \veps_t \text{~~~ (by \eqref{eq:eqJ})} \\
              &= \mC\trp\mJ\mC\mC\trp \vy_t + \mC\trp \veps_t \text{~~~ (because~} \mC\trp\mC = \mI\text{)} \\
              &= \mathbf{A}\vx_t + \mathbf{w}_t
\end{align*}
where we have defined the recurrence matrix $\mA = \mC\trp\mJ\mC$ and the latent dynamics noise $\mathbf{w}_t = \mC\trp \veps_t$ which follows $\vw_t \sim \Nrm(\vzero, \mQ)$ with $\mQ = \mC\trp\mP\mC$.
\\

Let us define $\vv_t = \vy_t - \mathbf{C}\vx_t = (\mI - \mathbf{C}\mathbf{C}\trp)\vy_t$. We need to determine the conditions under which $\vv_t$ is (i) normally distributed, and (ii) independent of $\vy_{t-1}$ and $\vx_t$. For this, we write:
\begin{align*}
    \mC\vx_t &= \mC\mA\vx_{t-1} + \mC\vw_{t-1} \\
             &= \mC\mC\trp\mJ\mC\vx_{t-1} + \mC\vw_{t-1} \\
             &= \mC\mC\trp\mJ\mC\mC\trp\vy_{t-1} + \mC\vw_{t-1} \\
             &= \mJ\vy_{t-1} + \mC\vw_{t-1}
\end{align*}
and hence:
\begin{align*}
    \vv_t &= \veps_{t-1} - \mC\vw_{t-1} \\
          &= (\mI - \mC\mC\trp)\veps_{t-1}
\end{align*}
which is independent of $\vy_{t-1}$ and has a marginal distribution $\vv_t \sim \Nrm(\vzero, \mR)$ with $\mR = \mP - \mC\mC\trp\mP\mC\mC\trp$, but is not in general independent of $\vw_{t-1}$. A sufficient and necessary condition for the independence of $\vw_{t-1}$ and $\vv_t$ is that the RNN noise covariance $\mathbf{P}$ has all its eigenvectors either aligned with or orthogonal to the subspace $\mathcal{F}$ (in this case, the covariance $\mR$ is degenerate and has $\mathcal{F}$ as a null space, which implies that observation noise is completely orthogonal to $\mathcal{F}$). If that is not the case, the reparametrization stays valid up to the fact that the observation noise $\vv_t$ and the latent dynamics noise $\vw_t$ can be correlated.
\\




\asection{Addition of input terms}
\label{sec:appD}
Let us consider an extension of both the latent LDS and the linear RNN models to take into account inputs. More specifically we will consider adding to both model classes an input under the form of a time-varying signal $u_t$ fed to the network through a constant set of input weights. In the latent LDS model, the input is fed directly to the latent variable and equations \eqref{eq:ldsx}-\eqref{eq:ldsy} become :
\begin{alignat}{3}
  \label{eq:ldsxi}
    \vx_{t} &=  \mA \vx_{t-1} \;  + \mB u_t \; &+& \; \vw_t,  \qquad & \vw_t &\sim
    \Nrm(\vzero,\mQ) \\
      \label{eq:ldsyi}
   \vy_{t} &= \mC \vx_t \;  &+ & \;  \vv_t, \qquad & \vv_t  &\sim \Nrm(\vzero,\mR),
 \end{alignat}
 
 The linear RNN equation \eqref{eq:rnn} becomes :
 \begin{equation}
  \label{eq:rnni}
  \vy_t = \mJ \vy_{t-1} + \mathbf{W}_{in} u_t + \veps_t, \qquad \veps_t \sim \Nrm(\vzero,\mP),
\end{equation}
so that we will represent by $\mB$ a low-dimensional input projection, and $\mathbf{W}_{in}$ a high-dimensional one.

For the LDS to RNN mapping, we can directly adapt the derivations of section \ref{mapping-lds-rnn-approx}, which lead to :
 \begin{equation}
   \label{eq:lati}
   \vy_{t+1} \mid \vy_t \sim \Nrm \left(\mC \mB u_t + \mJ_t \vy_t, \mP_t  \right)
 \end{equation}
 with the same expressions for $\mJ_t$ and $\mP_t$, given in equations \eqref{eq:ldstornn}-\eqref{eq:ldstornnvar}.
 
 \begin{sloppypar}
 For the RNN to LDS mapping, assuming again that $\mJ$ is low-rank and written as ${\mJ = \mM\mN\trp}$, we can define:
 \end{sloppypar}
 
 $$\vx_t = \mC\trp \vy_t$$
 
 where $\mC$ is a matrix whose columns form an orthonormal basis for the subspace $\mathcal{F}$ spanned by the columns of $\mM$, $\mN$ and $\mathbf{W}_{in}$. This latent vector then follows the \mbox{dynamics}:
 \begin{equation}
 \label{eq:rnnlati}
 \mathbf{x}_{t+1} = \mC\mJ\mC\trp\vx_t + \mC\trp\mathbf{W}_{in}u_t + \mC\trp\veps_t
 \end{equation}
 which corresponds to equation \eqref{eq:ldsxi}, and it is straightforward to show that it leads to equation \eqref{eq:ldsyi}, with the technical condition that the covariance of $\veps_t$ should have its eigenvectors aligned with the subspace $\mathcal{F}$ to avoid correlations between observation and recurrent noises.

\end{document}